# Enhanced cavity coupling to silicon monovacancies in 4-H Silicon Carbide using below bandgap laser irradiation and low temperature thermal annealing

*Mena N. Gadalla\*[1], Andrew S. Greenspon\*[1], Rodrick Kuate Defo[1,2], Xingyu Zhang[1], Evelyn L. Hu[1]*

[1]John A. Paulson School of Engineering and Applied Sciences, Harvard University, Cambridge, Massachusetts 02138, USA

[2]Department of Physics, Harvard University, Cambridge, Massachusetts 02138, USA

**Abstract**

*The negatively charged silicon monovacancy $V_{Si}^{(-)}$ in 4H-silicon carbide (SiC) is a spin-active point defect that has the potential to act as a qubit or quantum memory in solid-state quantum computation applications. Photonic crystal cavities (PCCs) can augment the optical emission of the $V_{Si}^{(-)}$, yet fine-tuning the defect-cavity interaction remains challenging. We report on two post-fabrication processes that result in enhancement of the $V1'$ optical emission from our 1-dimensional PCCs, indicating improved coupling between the ensemble of silicon vacancies and the PCC. One process involves below bandgap illumination at 785 nm and 532 nm wavelengths and above bandgap illumination at 325 nm, carried out at times ranging from a few minutes to several hours. The other process is thermal annealing at 100 °C, carried out over 20 minutes. Every process except above bandgap irradiation improves the defect-cavity coupling, manifested in augmented Purcell factor enhancement of the $V1'$ zero phonon line at 77K. The below bandgap laser process is attributed to a modification of charge states, changing the relative ratio of $V_{Si}^{(0)}$ ("dark state") to $V_{Si}^{(-)}$ ("bright state"), while the thermal annealing process may be explained by diffusion of carbon interstitials, $C_i$, that subsequently recombine with other defects to create additional $V_{Si}^{(-)}$s. Above bandgap radiation is proposed to initially convert $V_{Si}^{(0)}$ to $V_{Si}^{(-)}$, but also may lead to diffusion of $V_{Si}^{(-)}$ away from the probe area, resulting in an irreversible reduction of the optical signal. Observations of the PCC spectra allow insights into defect modifications and interactions within a controlled, designated volume and indicate pathways to improve defect-cavity interactions.*

**Introduction**

Silicon carbide (SiC) hosts a variety of color centers (point defects) with high brightness and long spin coherence times, even at room temperature.[1–3] It thus provides a promising platform for the realization of spin-based quantum technologies. As a technologically mature material with broad applications in high power electronics, LEDs, and microelectromechanical systems, large-area SiC wafers are available in a variety of polytypes, facilitating both fundamental studies as well as providing pathways for technological scale-up. Appropriate integration of SiC color centers within high quality optical cavities is an important enabler for the development of quantum information technologies in this material: resonantly tuned cavities can enhance the optical emission of color centers, enabling substantial increase of the color center emission into the zero-phonon line (ZPL), rather than to a phonon-broadened background. Although cavity-color center emission enhancement has been demonstrated for color centers in 4H-SiC, [4–7] an outstanding challenge relates to the optimal placement of the color centers within the optical cavity, and post-fabrication fine-tuning the defect-cavity interaction. This paper describes two processes to fine-tune the defect-cavity interaction, enabling the optimal placement of the color centers within the cavity.

We examine negatively charged silicon monovacancies $V_{Si}^{(-)}$ in 4H-SiC, integrated within 1D photonic crystal cavities (PCC), and observe enhanced emission from the cavities, subject to either laser illumination at two different below-bandgap wavelengths, or low temperature thermal annealing at 100 °C. We also subject PCCs to above-bandgap laser illumination and observe a transient enhanced emission followed by a longer term decay in the optical signal. The different thermal annealing and laser illumination processes produce distinctive changes in the PL spectra at both room temperature and at 77 K. At 77 K, the zero-phonon lines (ZPLs) are visible for the $V_{Si}^{(-)}$ emission, resulting from defects positioned at two different lattice sites, the $V1$ at the hexgonal (h) site and the $V2$ at the cubic (k) site.[8] The $V1$ center has a shorter wavelength companion ZPL, termed $V1'$ (859 nm) resulting from a transition to an electronic excited state of slightly higher energy than the excited state corresponding to the $V1$ transition (861 nm).[9] More recent computational modeling suggests that the $V1'$ ZPL is due to a polaronic excited state with mixed electronic character between two pure electronic states due to interactions with atomic vibrational excited states in the SiC lattice.[10] We chose to work with $V1'$ for two reasons: 1) The alignment of the $V1'$ emission dipole relative to the optical cavity's electric field polarization results in stronger coupling to the cavity field, and 2) the uncoupled $V1'$ ZPL intensity is brighter than that of the $V1$ at 77 K.

**Measurements**
Similar to techniques described in our previous work, we employ a 1D nanobeam photonic crystal cavity (PCC) design that provides a high theoretical quality factor, $Q$, with a small mode volume, V.[11,12] Color centers were introduced into the nanobeam PCCs post-fabrication through carbon ion implantation (*Cutting Edge Ions Inc.*). More details regarding the fabrication and implantation are provided in the supplementary information. These PCCs were designed to be resonant in frequency with $V1$ and $V1'$; however, ion implantation creates a diversity of defects other than the color center of choice,[13] for example carbon vacancies ($V_C$), Si-C divacancies (*VV*), anti-site defects (C on a Si site $C_S$, or Si on a C site $Si_C$), as well as Si and C interstitials ($Si_i$, $C_i$, respectively). This ambient defect environment is important in understanding our experimental observations. Without any post-implantation annealing, a typical room temperature photoluminescence (PL) spectrum appears as shown in Figure 1. Analysis was carried out on several PCC structures. In general, quality factors ranged from 1500-3000, and un-tuned, room temperature spectra showed prominent cavity mode PL peaks.

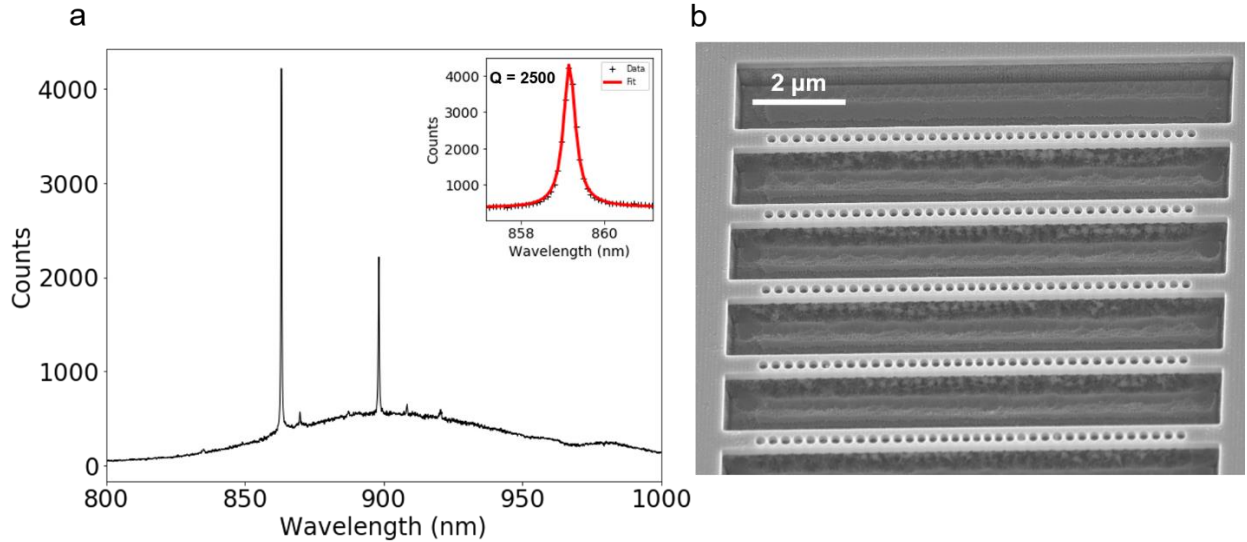

*Figure 1: (a) A typical photoluminescence spectrum from a SiC nanobeam. The broad background reflects emission from optically active impurities in the beam. The narrow peaks are the modes of the cavity coupled to ensembles of $V_{Si}^{(-)}$s. The shortest wavelength mode is the fundamental mode of the cavity which is designed to be resonant with the ZPL of $V1$ and $V1'$ transitions at 859 and 861 nm respectively. Inset: A Lorentzian fit of the fundamental cavity mode. (b) SEM micrograph of suspended nanobeam PCCs with the cavity mode electric field polarization parallel to the beam direction.*

Below-band gap laser illumination was performed on a LabRam Evolution Horiba Multiline Raman Spectrometer using either a continuous wave (CW) 785 nm red laser at 50 mW or a CW 532 nm green laser at 30 mW, using a 100x objective. Above-band gap illumination was also carried out in the same system, using a CW laser operating at 325 nm and 4 mW, with a 40x objective. The 100x objective lens with ~800 nm beam diameter was positioned at the center of the nanobeam where the cavity region is located and focused on the $V_{Si}^{(-)}$ ensemble in the z-direction (location of maximum PL signal). Illumination was carried out for various times, ranging from a few minutes to several hours. Thermal annealing was performed on a hot plate. After the hotplate reading stabilized at 100 $°C$, the sample was added and annealed for 20 minutes.

Room-temperature PL measurements were carried out using micro-PL (uPL) confocal spectroscopy performed on a LabRam Evolution Horiba Multiline Raman Spectrometer, under a 100x objective. The illuminating radiation was a CW 785 nm laser at 500 uW power, a power far lower than the powers used for the laser illumination treatments.

A more definitive assessment of the laser and thermal treatments on the samples can be gained through low temperature (77 K) measurements, taken as the cavities were tuned into resonance with the $V1'$ ZPL at 859 nm, allowing determination of the Purcell enhancements. Tuning was carried out using carbon dioxide ($CO_2$) gas condensation onto the cavities. These measurements were performed with a Janis flow-through cryostat and a home-built confocal microscope using a pulsed 760 nm laser (Mira 900, 76 MHz repetition rate, <200 fs autocorrelation, pumped by a Verdi-V10 532 nm laser at 10 W) at 140 uW time-averaged power and a 40x long-working distance objective with aberration correction collar.

## Results

*General features of steady state changes in the PL spectra.*

Before providing more detailed discussions of the various annealing processes, an initial overview of the raw PL data presented in Figure 2a can illustrate the principal changes in the optical signals that result from thermal annealing and laser illumination. For thermal annealing, below-bandgap illumination, and the initial stages of UV laser illumination, we observe an enhancement in the optical signal (in both the background and the optical cavity mode intensity. We will refer to the latter as the $V_{Si}^{(-)}$ peak). Red and green laser illumination produce an optical signal that increases with time, reaching a limiting value under prolonged illumination time. UV illumination, however, induces an initial increase in the optical signal, followed by a continuous decrease with prolonged illumination. With the exception of long-time UV-illuminated PCCs, no changes in the quality factor (as measured by linewidth of the cavity mode) were observed for the laser-illuminated or thermally-annealed nanobeam cavities.

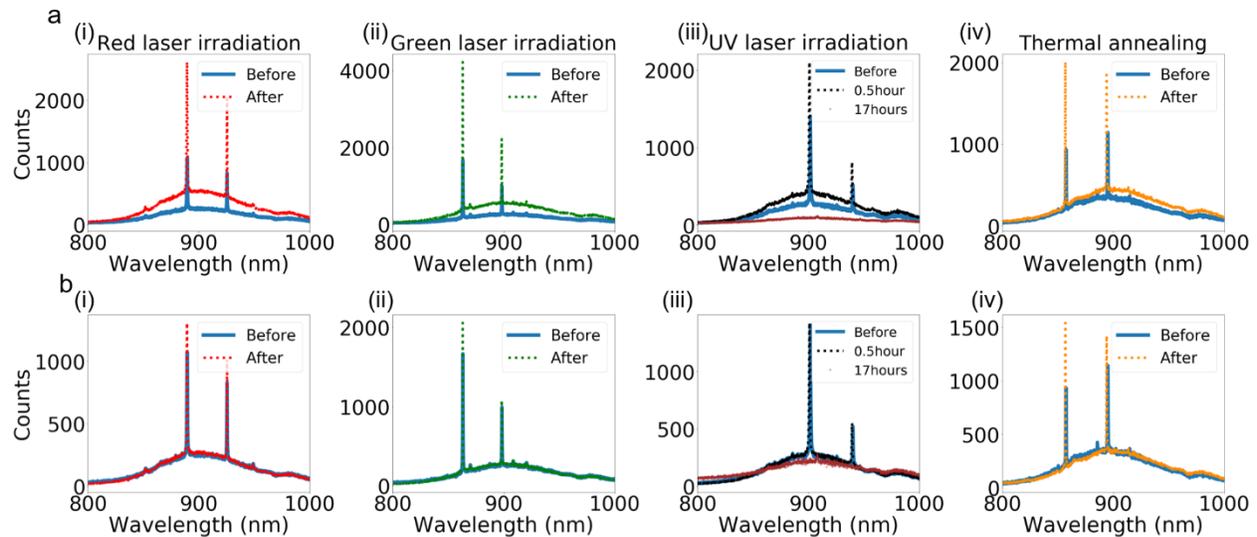

Figure 2: Steady state effect of laser illumination and low temperature thermal annealing on PCC optical spectra: (a) Initial PL spectra for a cavity before (blue) and after (i) red (ii) green and (iii) UV illumination and (iv) thermal annealing at 100 °C. For red and green laser irradiation, the illumination was performed until saturation at the maximum enhancement in the optical signal was reached. (b) The corresponding PL spectra where the backgrounds of the initial spectra have been multiplied by a numerical factor to overlap with the backgrounds of the after-process spectra. (Overlap is only approximate for prolonged UV illumination). The increase in the $V_{Si}^{(-)}$ signal is then clearly apparent.

Figures 2a (i)-(iv) also delineate the distinctive differences in the *backgrounds* of the spectra subjected to the various processes. For PCCs illuminated by lasers at red and green wavelengths, there is a steady increase in background count with increased illumination time. Like the peak intensity of the cavity modes, the backgrounds reach limiting values. For the UV-illuminated PCCs, the background initially increases, and then decreases with prolonged illumination. There is a far lower increase in background for the thermally annealed PCCs.

In general, the relative shape (envelope) of the background does not appear to change after illumination or thermal annealing, so that a simple multiplicative factor, computed as the integration of the background without cavity modes, works well to provide normalized values of the optical signal from the cavity, as illustrated in Figure 2b. The different background behaviors are discussed in greater detail in the section below and will be important in further understanding and interpreting the results of the laser and thermal treatments.

*Below bandgap illumination*

Previous studies have focused on changes in the total optical signal from $V_{Si}^{(-)}$ defects in the bulk due to laser illumination or thermal annealing.[14,15] In this study, we take advantage of the interaction between the PCC and silicon vacancies to study the transient response of the defect to thermal annealing or laser illumination at incremental time steps. Figure 3 shows the evolution of the $V_{Si}^{-1}$-PCC spectra as a function of total laser illumination time for two representative PCCs using illumination at 785 nm (50 mW power) (Figure 3a (i)), and 532 nm (30 mW power) (Figure 3b (i)). Figures 3a (ii) and 3b (ii) show the change in normalized intensity of the fundamental cavity mode with increased illumination time. The insets provide greater detail of the fitted peaks, and the corresponding quality factors. Details on the fitting of the peaks can be found in the supplementary information. The percent change we observe in the PCCs varies from 5 to 25% but averages ~20-25% for the twelve cavities analyzed, six with 532 nm and six with 785 nm irradiation (Table 1). Additionally, Figures 3a (iii) and 3b (iii) show the percentage increase in the integrated background with increased illumination time. The percent change in backgrounds varies from 20 to 100% depending on the PCC measured, averaging ~70-90% over all nanobeams studied (Table 1). Although Figure 3 shows data for particular cavity-defect samples, these cavities illustrate several features that generally pertain to all samples we studied, subjected to 785 and 532 nm illumination:

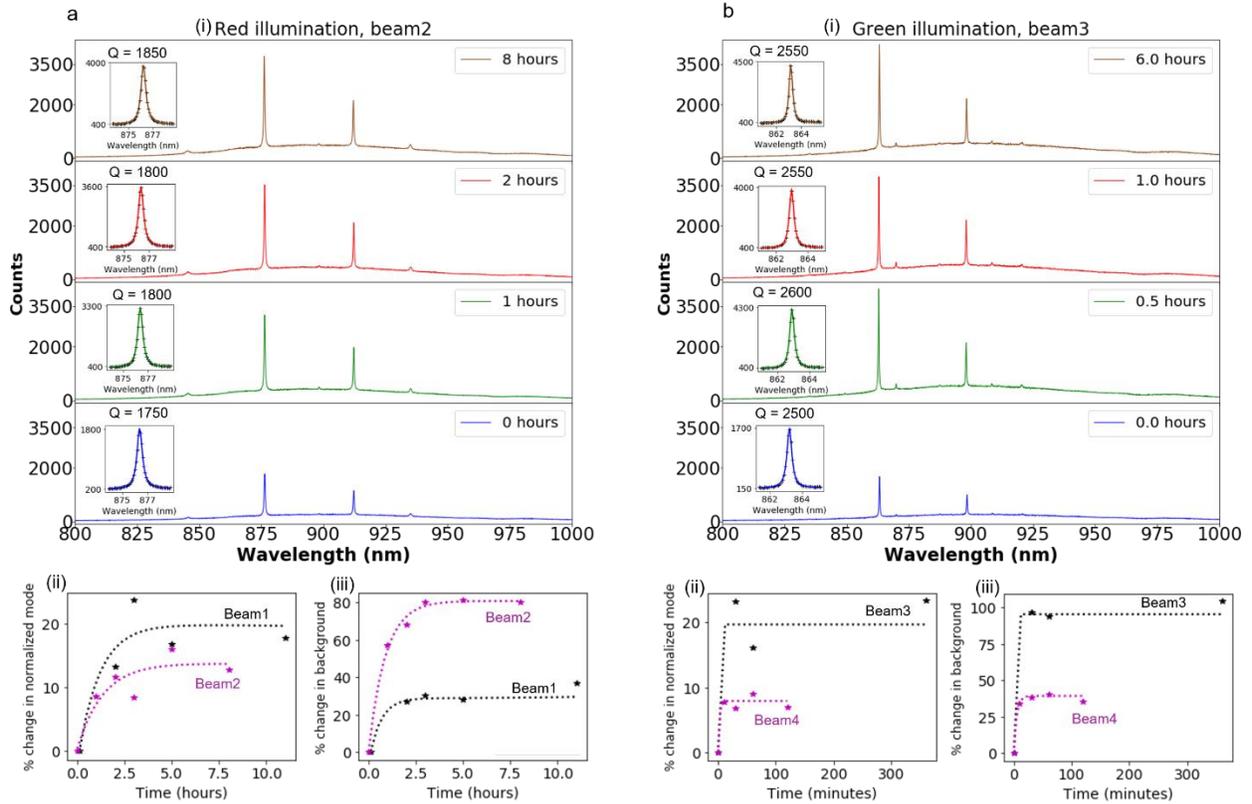

*Figure 3: Evolution of PL spectra of cavity with increased laser illumination time, showing significant increase in the optical signal of the mode for laser illumination at (a) 50 mW, 785 nm and (b) 30 mW, 532 nm: i) Time sequence of successive PL spectra for a given PCC after each illumination step. Transient increase of (ii) the normalized intensity of the fundamental mode, and (iii) the integrated background for two different PCCs. Red and green illumination data show a saturation time of 4-5 hours and 20 minutes respectively.*

1. Within experimental error, there is no change in the peak positions, nor in the widths of the peaks, implying that there is no change in the corresponding Q of the modes. (See the insets of Figure 3).
2. There is a variation in the relative increase in peak height among the different defect-PCCs analyzed. For a given illumination wavelength, however, the saturation time is about the same for different PCCs.
3. There is a more rapid saturation to the maximum PL peak intensity with 532 nm illumination compared to 785 nm illumination.
4. For both red and green laser illumination, the saturation times for the change in normalized cavity mode intensity and the change in background are the same for a given PCC.

*Above bandgap illumination*

Laser illumination was also carried out using 325 nm (3.8 eV) at 4 mW power, with a 40x NUV objective (Figure 4). Initially, the laser illumination at 325 nm led to an increase in the optical signal (both normalized intensity of cavity mode and background) for illumination times shorter than an hour. However, in contrast to the below-bandgap illumination process, prolonged UV illumination led to a decrease in the optical signal, and in some cases, the complete disappearance of the cavity modes.

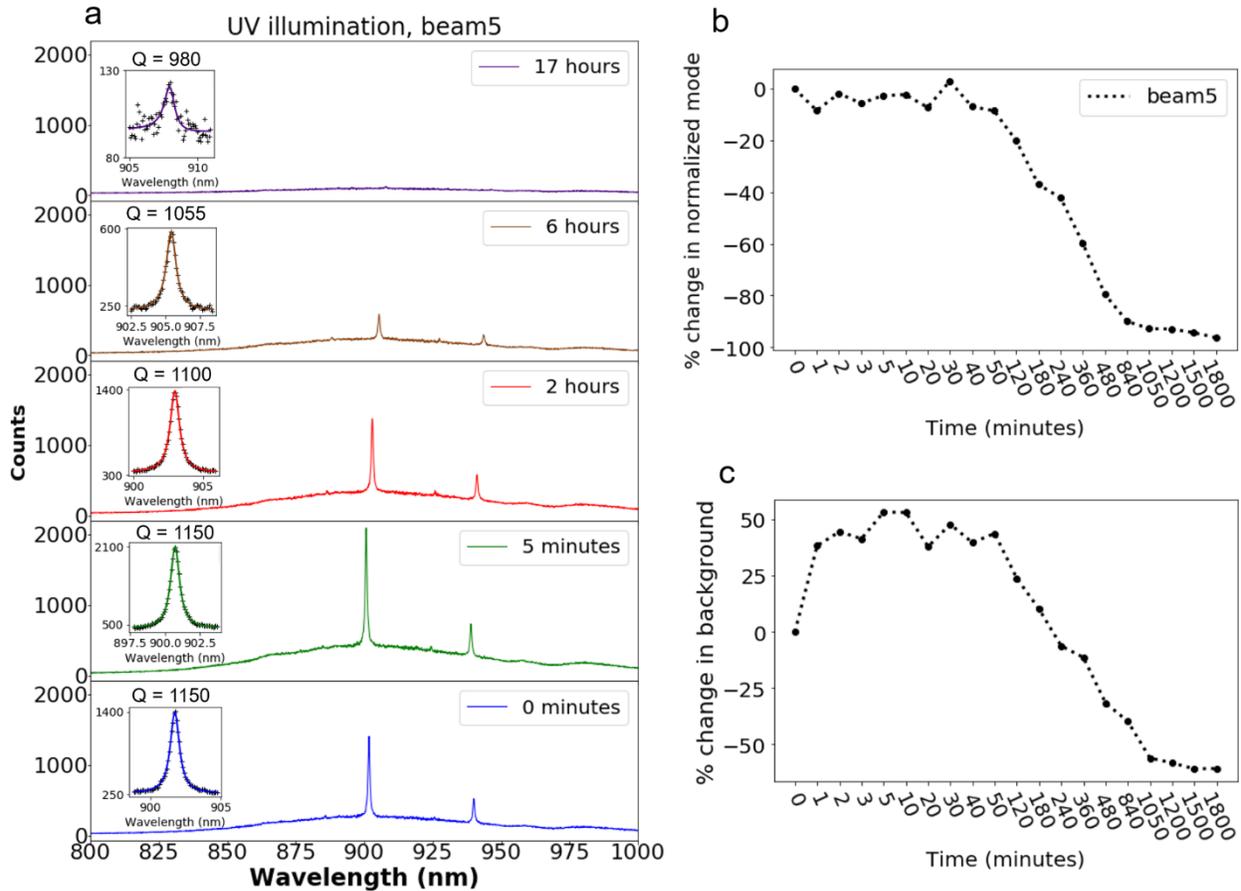

*Figure 4: (a) Evolution of PL spectra of a cavity for increased UV illumination times. (b) Transient evolution of the change in the normalized intensity of the fundamental cavity mode in the PL spectra, showing a significant decrease in PL of the cavity peak with prolonged illumination time. (c) Transient evolution of the change in the integrated background. Initially, the background increases significantly. Over hours-long illumination times, we observe a significant decrease in the background.*

*Thermal annealing*

Similar to the results found with below bandgap laser annealing of the cavities, thermal annealing at 100 °C for 20 minutes results in increased PL intensity of the fundamental cavity mode. A distinctive feature of these spectra, compared to the spectra of laser annealed samples, is the almost negligible increase in background, while achieving substantial enhancement of the cavity mode PL emission.

To complement the room-temperature data, we carried out cavity tuning measurements at 77 K, using carbon dioxide ($CO_2$) gas condensation onto the cavities to tune the modes into resonance with the $V1'$ ZPL at 859 nm. The results confirmed the room-temperature results, with as high as a three-fold increase in Purcell factors resulting from the below bandgap laser illumination and low temperature thermal annealing (Supplementary Info). Table 1 provides a summary of the general features of the different processing conditions.

*Table 1: Summary of room temperature PL changes over time due to laser illumination and thermal annealing. The normalized cavity mode peak intensity was computed as the integration of the mode after scale correcting the spectra, followed by background subtraction. The data were collected from six beams for each process.*

|  | Change in normalized fundamental cavity mode intensity | Change in Background | $\tau_{sat}$ |
| --- | --- | --- | --- |
| 785 nm, 50 mW | 1.20±0.06 | 1.72±0.23 | ~4-5 hours |
| 532 nm, 30 mW | 1.25±0.14 | 1.90±0.12 | ~20 mins |
| 325 nm, 4 mW | 0.185±0.015 | 1.00±0.06 | N/A |
| 100 °C, 20 minutes | 2.10±0.25 | 1.25±0.16 | N/A |

**Discussion**

The emission of defects in PCCs provides a distinctive *nanoscope* into the atomic-scale interactions of the defects within that cavity, and a method to understand the laser illumination and annealing experiments described above. The PCCs have been engineered and fabricated to be resonant with a particular defect ($V_{Si}^{-1}$) emission frequency (859 nm) and emission dipole ($V_1'$), with the cavity mediating an enhanced spontaneous emission of the chosen emitter, resulting in a *Purcell Enhancement, F:*

$$F = \frac{3}{4\pi^2} \xi \frac{Q}{V} \left(\frac{\lambda}{n}\right)^3$$

$$\xi = \frac{\omega_c^2}{4Q^2(\omega_c - \omega_0)^2 + \omega_c^2} \left(\frac{|\mu.E|}{|\mu||E_{max}|}\right)^2$$

$Q$ is the quality factor of the cavity, $V$ is the mode volume, $n$ is the refractive index of the material, and $\lambda$ the wavelength of the cavity mode. The factor $\xi$ accounts for the spectral and spatial overlap of the cavity field with the emitter being enhanced, as well as the alignment of the cavity field and the emitter's optical dipole. Thus, a maximal value of F results from a cavity resonance mode ($\omega_c$) tuned to the emitter frequency ($\omega_0$), with the emission dipole of the emitter parallel to the cavity's electric field polarization ($\mu \cdot E = |\mu||E|$), and with maximal spatial overlap with the modal field ($E = E_{max}$).

Thus, these studies provide insight on the distribution of an ensemble of $V_{Si}^{(-)}$ defects within a volume of approximately (0.1 μm)$^3$, the approximate modal volume of these PCCs. Given the $10^{12}$ cm$^{-2}$ implantation dose used to form the defects, one can estimate that approximately 100 ions are stopped within that volume, $V$. Using published estimates of the efficiency of creation of a silicon vacancy by a bombarding carbon ion to be as high as 20%,[16] we approximate ~ 20 $V_{Si}^{-1}$s spatially distributed within the volume that is probed by our cavity.

Since the illumination and annealing treatments described above *do not* result in any statistically significant change in the values of $Q$ of our cavities (ranging from 1500 – 3000), nor in a shift in the modal frequencies, it is reasonable to assume that the enhancements observed are a result of the better spatial overlap of the $V_{Si}^{(-)}$ s with the modal fields. A simple, first-order assumption is that the thermal or laser processing (a) might promote the diffusion of $V_{Si}^{(-)}$ s into greater spatial overlap with the cavity mode, or (b) that the treatments create additional $V_{Si}^{(-)}$ within the modal volume, or (c) perhaps a combination of (a) and (b).

*Enhancement through laser illumination*

The charge states of defects influence both spin behavior as well as luminescence efficiency (i.e., whether the defect is "dark" or "bright"). For example, the negatively charged monovacancy in 4H-SiC, $V_{Si}^{(-)}$ is "bright", while the neutral state, $V_{Si}^{(0)}$, is "dark".[15] The neutral divacancy in 4H-SiC, $VV^0$ is bright, while $VV^-$ is dark.[17] Accordingly, a number of recent studies have utilized applied voltages or laser illumination of defects to better understand and controllably alter the charge states of those defects, as a means of enhancing optical read-out of defect states.[15,18–20] Such studies build upon representations of the charged defect states within the bandgap of 4H-SiC, as indicated in Figure 5, which shows various charge states for the carbon vacancy, $V_C$, the divacancy, $VV$, as well as the silicon vacancy, $V_{Si}$. We understand the over-simplification of the defects that actually may be present in our PCCs; nevertheless, this simple representation helps to qualitatively explain some key features of our data.

The interpretation of our below bandgap laser annealing data must take into account the following experimental observations: (a) although there are variations in total PL enhancement among all defect-cavity systems, the maximum normalized enhancement observed averages a factor of 1.2-1.25, regardless of whether red or green illumination was used, and (b) the times required to reach the maximum value of PL were dramatically shorter for the green annealing treatments.

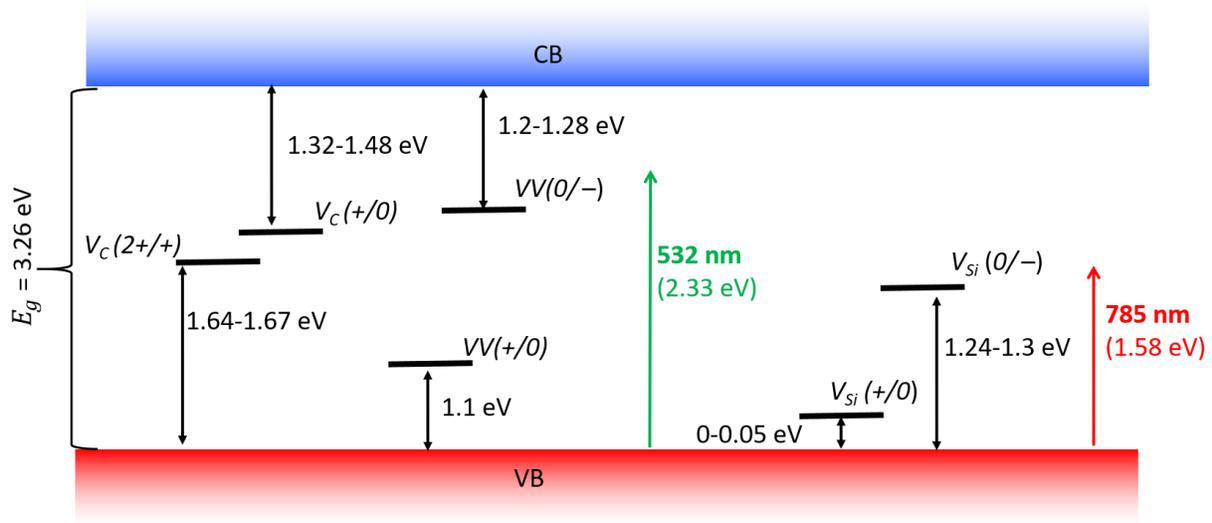

Figure 5. Location of different defect states within the bandgap of SiC, along with the energies of 785 nm and 532 nm laser excitation. The green laser allows excitation of an electron from the valence band to divacancy and carbon vacancy defect states, followed by excitation of the electron to the conduction band, where it can recombine with a silicon vacancy to generate $V_{Si}^{-1}$. The red laser does not have enough energy to access some of these defect states directly. Estimates of energy levels from [Hornos 2011]; [Gordon 2015]; Magnusson 2018].[17,21,22]

An initial focus on the charge states of the silicon monovacancy alone indicates that both red and green lasers have sufficient energy to promote electrons from the valence band to form the negatively charged ("bright") silicon vacancy state. But Wolfowicz's photo-pumping experiments on bulk 4H-SiC revealed a strong relationship between $VV$ and $V_{Si}$, where excitation at energies above the $VV^-$ transition (~ 1.3 eV) can release electrons to the conduction band that may subsequently be captured by a neutral silicon vacancy to form $V_{Si}^{(-)}$.[15] Both red and green lasers have enough energy to cause this transition. However, only illumination by the green laser (2.33 eV) has the ability to promote valence band electrons to the $VV^-$ state. Subsequent photo-ionization of the $VV^-$ states can then promote electrons to the conduction band that may subsequently be captured by a neutral silicon vacancy to form $V_{Si}^{(-)}$. Proximal carbon vacancies may similarly play a role in more rapidly forming $V_{Si}^{(-)}$s: illumination at 2.33 eV may promote valence band electrons to the $V_C^{(0)}$ state, with subsequent photo-ionization creating additional electrons in the conduction band that may help form $V_{Si}^{(-)}$. To summarize, both red and green annealing processes achieve similar levels of $V_{Si}^-$ density at saturation, presumably a function of the number of silicon vacancies (charged or uncharged) present within the modal volume. The higher-energy green laser, however, in interacting with other defects in that volume, provides more pathways to producing $V_{Si}^{(-)}$s, thereby more quickly (minutes) reaching the limiting value of $V_{Si}^{(-)}$-associated PL enhancement rather than the hours needed to achieve the charge conversion using the red laser. Laser-induced charge-state conversion, and the production of a larger number of "bright" defects throughout the PCC can also account for the increase in the backgrounds with a similar saturation and time-dependence as the increase in the $V_{Si}^{(-)}$ peak.

The model we have proposed is only qualitative at the moment and requires further experiments and detailed calculations of various photo-ionization and capture rates. However, these initial experiments suggest that we may gain some important insights into the relative percentage of neutral to charged silicon vacancies, and also about the proximal distributions of carbon vacancies and divacancies within the volume pertaining to our cavities.

*Reduced optical signal observed with above bandgap illumination*

Wolfowicz et al observed a dramatic decrease in the $V_{Si}^{(-)}$ intensity after illumination by above-bandgap (365 nm, 3.38 eV) illumination. This appears to have similarities with the results of our above-bandgap experiments. However, Wolfowicz reported that subsequent illumination at either 978 nm or 780 nm resulted in the restoration of the $V_{Si}^{(-)}$ state, whereas for our PCC samples, subsequent red or green illumination for periods as long as 24 hours failed to restore the $V_{Si}^{(-)}$ luminescence peak. Therefore, we believe that different mechanisms are responsible for the reduction of $V_{Si}^{(-)}$ PL in our structures.

Wolfowicz et al attribute their loss of $V_{Si}^{(-)}$ signal from UV excitation to intermediary nitrogen defect levels, close in energy to the conduction band, that trap the electrons and reduce electron capture by $V_{Si}^{(0)}$. In this case, subsequent lower-energy laser illumination releases those trapped electrons, forming additional $V_{Si}^{(-)}$.

Our PCCs are intrinsically-doped, so our experiments may not include those defect levels. In our experiments, above-bandgap illumination at 325 nm (3.8 eV) results in an initial increase in the $V_{Si}^{(-)}$ PL signal for time periods lasting about 60 minutes. There is an accompanying increase in the background signal as well. However, prolonged exposure to UV illumination results in a decrease in the optical signal, and eventually the disappearance of the optical emission from the $V_{Si}^{(-)}$ (Figure 4). There is an accompanying decrease in the spectral background. It is likely that the initial increase in optical signal and background indicates charge state conversion of $V_{Si}^{(0)}$ to $V_{Si}^{(-)}$, resulting from electron-hole pair generation in the valence and conduction bands, followed by $V_{Si}^{(0)}$ capture of electrons from the conduction band. The background impurity levels of our PCCs are likely different from the materials used in the Wolfowicz experiments, which may explain why we initially observe an increase in $V_{Si}^{(-)}$ signal.

The *permanent* loss of the optical signal for prolonged UV illumination (> 60 min), as well as the accompanying reduction of the background, raises the possibility of diffusive motion of $V_{Si}^{(-)}$ out of the region of maximum cavity field, and eventual encounter with the etched sidewalls of the cavity. The percent decrease in background is less than that of the normalized mode intensity. This is consistent with the diffusive movement of the $V_{Si}^{(-)}$ outside of the principal cavity volume, while the residual impurities in the PCC form the observed background. Many more detailed experiments will be needed to more fully explicate our data, but we note the possibility of recombination-enhanced defect motion,[23] where above bandgap excitation can lead to non-radiative recombination by multi-phonon emission, localized at the defect, and lowering the barrier to defect diffusion. Indeed, such behavior has been observed in 4H-SiC.[24]

Fuchs et al. focused 785 nm laser light of powers ranging from 0.5 mW to 3.5 mW on heavily neutron-irradiated ($5x10^{17} cm^{-2}$) 4H-SiC bulk material for 120 minutes. They found an increase in the $V_{Si}^{(-)}$ PL intensity as a function of laser power, observing a 25% enhancement of signal for the 3.5 mW illumination. They postulated two possible reasons for the enhancement, laser-induced change of charge state or local heating by the laser, although they did not make a final attribution of mechanism to observed enhancement. We believe that laser powers used in all our illumination experiments are insufficient to induce local heating that could cause diffusive motion of defects. Using the formalism from research into laser heating,[25–27] at room temperature, we find a steady-state largest increase in temperature of about 0.8 K for 325 nm illumination at 4 mW, 0.02 K for 532 nm at 30 mW, and 0.008 K for 785 nm at 50 mW.[28]

*Thermal annealing*

Fuchs et al carried out thermal annealing experiments on their heavily neutron-irradiated samples, at temperatures ranging from $125^oC$ to $700^oC$ for 90 minutes.[14] They observed negligible enhancement of $V_{Si}^{(-)}$ PL for temperatures below ~150 ºC, followed by a monotonic increase in PL signal up to 600 °C. The PL enhancement reaches a maximum enhancement of 5.5, and then dramatically decreases with further increase of temperature. Fuchs estimate that their initial $V_{Si}^{(-)}$ concentration corresponds to $7x10^{15} cm^{-3}$, distributed almost homogeneously in a bulk sample. They attribute the temperature-induced PL enhancement to recombination of various defects to create additional $V_{Si}^{(-)}$s, as well as some changes of charge state.

The PCCs allow us to make sensitive measurements of $V_{Si}^{(-)}$ luminescence within a well-defined, limited volume, and we do observe an enhancement in PL for annealing at 100 °$C$ for only 20 minutes. In considering possible mechanisms, calculations[29] suggest that carbon interstitials, $C_i$s with the relatively low activation energies of ~ 0.95 eV can play a principal role in the change in distribution of defects, and hence the change in PL. First-principle calculations suggest that $C_i$ exists primarily in the neutral charged state in intrinsically doped SiC,[29] which is the area our defects reside in. Based on an average diffusion barrier of 0.95 eV, we calculate the diffusion constant of $C_i$ to be ~5.5 $x\ 10^{-19} m^2/s$ at $100^oC$, corresponding to a diffusion length of 26 nm over 20 minutes. The diffusion constant is $1.9\ x\ 10^{-22} m^2/s$ at $20^oC$, corresponding to a diffusion length of 0.5 nm over 20 minutes. Therefore at room temperature, the $C_i$ is essentially frozen in place (may move one lattice site after 20 minutes), whereas at $100^oC$, it can move ~100 lattice sites and combine with nearby defects.

Mobile neutral $C_i$ can give rise to the following changes: (a) recombination with $V_C$, restoring a C within the SiC lattice. This should produce no change in the $V_{Si}^{(-)}$ PL from the cavity; (b) recombination with $a$ $V_{Si}$ to form a C anti-site defect, decreasing the number of $V_{Si}$s and hence decreasing the PL intensity from the cavity; (c) recombination with a neutral divacancy, forming $V_{Si}^{(0)}$, leading to no change in PL intensity; or (d) recombination with a negatively charged divacancy, forming a $V_{Si}^{(-)}$, leading to an increase in the PL of the cavity. Therefore mechanism (d) could be responsible for the enhanced PL observed as a result of low-temperature thermal annealing. In addition, mechanism (c) followed by a charge-conversion process should also result in PL enhancement (these studies will be described in a later paper).

Thermally annealed samples demonstrate a far lower increase in background than do the below-bandgap laser-illuminated samples. Some experiments have been initiated to further explore this effect and will be reported in a later paper. The red and green laser treatments appear to stabilize the PL enhancement, in a way not observed for the thermal treatment, and this may provide some guidance in further understanding the differences in background growth.

As is true for our laser irradiation treatments, these thermal annealing experiments at the moment raise possibilities but do not provide definitive answers. But both sets of experiments indicate our ability to augment the PL signature of the $V_{Si}^{(-)}$, and both sets of experiments help us begin to map out relative positions of charged and neutral-state silicon vacancies, divacancies, and ultimately, carbon vacancies.

*Conclusion*

Using PCCs as an effective nanoscope, we have developed a better understanding of the effects of different frequencies of laser irradiation and low temperature thermal annealing on the charge state and motion of point defects in silicon carbide. In particular, we have analyzed how these processing conditions influence defect-cavity coupling for the negatively charged silicon vacancy, a potential qubit for quantum memories.

Below-bandgap laser annealing in our samples leads to charge state conversion, and the increase in PL signal results from the conversion of $V_{Si}^{(0)}$ to $V_{Si}^{(-)}$, where green laser illumination reaches conversion saturation more rapidly than red due to additional pathways available to move electrons from the valence band to the conduction band via other defect states. Above-bandgap laser illumination initially appears to similarly lead to charge state conversion with an increase of PL signal. However, longer-duration illumination produces a dramatic decrease in the PL, indicating a different underlying physical mechanism that warrants further exploration. By contrast, we believe that thermal annealing at 100 °C resulted in enhanced PL through a very different mechanism: diffusion of $C_i$s and recombination with $VV^-$ to form additional $V_{Si}^{(-)}$s.

There is a natural stochastic variation in defect type and distribution within the modal region of our PCCs. Even so, these initial experiments have shown our ability to augment the PL signature of $V_{Si}^{(-)}$ from the cavity through charge-state control and defect recombination mechanisms. Better understanding can result from finer granularity in changes of laser wavelengths and power, annealing temperatures and times, and by varying the initial concentration of defects within the cavities.

Our initial results suggest that a combination of thermal annealing and below-bandgap laser irradiation can improve defect-cavity coupling for the $V_{Si}^{(-)}$, which will be essential for possible quantum memory devices. More broadly, the PCC nanoscope provides a constrained, nanoscale volume for probing defect behavior that yields important insights into optimizing defect-cavity interactions for a wide range of applications.

**Supporting Information**

Supporting Information is available online or from the author.

**Authors Contribution**

M. N. G. performed cavity simulations, nano-structure fabrication, annealing and irradiation experiments, and data collection via optical measurements at room and cryogenic temperatures. A.S.G. helped with nano-structure fabrication. M.N.G. and A.S.G set up optical systems and cryogenic setups for data collection. R. K-D. ran Density Functional Theory simulations to characterize defect diffusion mechanisms in 4H-SiC. X. Z. performed cavity design and initial fabrication experiments. M. N. G., A.S.G, R. K-D and E.L.H. analyzed the data and co-wrote the manuscript.

**Acknowledgements**

M. G. acknowledges support from National Science Foundation RAISE TAQS grant number 1839164-PHY and EAGER grant number ECCS-1748106. A.S.G. acknowledges support from the STC Center for Integrated Quantum Materials under National Science Foundation Grant No. DMR – 1231319. This work was performed in part at the Harvard University Center for Nanoscale Systems (CNS), a member of the National Nanotechnology Coordinated Infrastructure Network (NNCI), which is supported by the National Science Foundation under NSF ECCS award No. 1541959. This work used computational resources of the Extreme Science and Engineering Discovery Environment (XSEDE), which is supported by National Science Foundation Grant Number ACI-1548562, [ J. Towns, T. Cockerill, M. Dahan, I. Foster, K. Gaither, A. Grimshaw, V. Hazlewood, S. Lathrop, D. Lifka, G. D. Peterson, R. Roskies, J. R. Scott, and N. Wilkins-Diehr, Computing in Science & Engineering 16, 62 (2014)] on Stampede2 at TACC through allocation TG-DMR120073, and of the National Energy Research Scientific Computing Center (NERSC), a U.S. Department of Energy Office of Science User Facility operated under Contract No. DE-AC02-05CH11231.

**Conflict of Interest:** The authors declare no competing financial interest.

**Supplementary Information For "Enhanced cavity coupling to silicon monovacancies in 4-H Silicon Carbide using below bandgap laser irradiation and low temperature thermal annealing"**

*Mena N. Gadalla*[1], *Andrew S. Greenspon*[1], *Rodrick Kuate Defo*[1,2], *Xingyu Zhang*[1], *Evelyn L. Hu*[1]

[1]John A. Paulson School of Engineering and Applied Sciences, Harvard University, Cambridge, Massachusetts 02138, USA

[2]Department of Physics, Harvard University, Cambridge, Massachusetts 02138, USA

*Silicon carbide defect implantation conditions*

Starting materials were 4H-SiC p-i-p n wafers provided by Norstel AB. The thicknesses of the epitaxial layers are p-i-p (100, 200, 100 nm) layer. Then n-type substrate is nitrogen doped at $10^{19}/cm^3$. The p-type layers are aluminum doped at $10^{19}/cm^3$. Details of the PCC fabrication can be found in [Bracher 2017].[1] The photoelectrochemical (PEC) etch we used to remove the n-type layer and undercut our beams often partially removed the lower p-type layer as well. For the sample under study in this publication, the final thickness of the suspended nanobeam PCC is 300 nm (100 nm p-layer on top of 200 nm i-layer). After nanobeam fabrication, silicon monovacancy centers were created by implantation (*CuttingEdge Ions, LLC*) of $^{12}C$ ions at a dose of $10^{12}/cm^2$ and an energy of 170 keV, with a sample tilt angle of 7 degrees from horizontal. In the vertical dimension, the cavity field profile reaches its maximum value in the middle of the cavity, which is 150 nm below the surface for our 300 nm device layer. SRIM simulations show that our implantation dose gives a maximum defect creation at ~250 nm, well positioned in the intrinsic doped layer.

*Method for peak fitting and calculating intensity of cavity modes*

Fitting and calculation of the intensity of each cavity mode was performed as follows: A linear fit was applied to the background signal around the cavity mode – this linear fit was then corrected to a flat background and removed. The cavity mode was then fit to a standard Lorentzian lineshape to extract the quality factor and the intensity of the cavity mode was calculated as the integral under the curve using the composite trapezoidal rule. The transient response in Figure 3 in the manuscript for red and green illumination fit using the function, $c_0 + c_1 t^3 + c_2 e^{-c_3 t}$ where t represents time and $c_0, c_1, c_2,$ and $c_3$ are the fitting constants.

*Scale factors to normalize changes in the intensity of the cavity mode*

To better understand the effect of laser illumination and thermal annealing on the coupling between the photonic crystal cavity and the defects, we must account for changes in the background photoluminescence (PL) signal after any given processing step. Therefore, for every PL spectrum, we integrate under the whole curve and subtract the integrated intensity from each cavity mode to determine the integrated signal solely due to the PL background. We then divide the integrated background after processing by the integrated background before processing to obtain a scale factor. For below bandgap illumination, thermal annealing and the initial stages of UV illumination, the relative shape (envelope) of the background does not change. As such, using this scale factor, we normalize each before and after spectrum to obtain an accurate value for the enhancement in the intensity of the cavity modes. This multiplicative factor also tells us the percent change in background photoluminescence from defects in the ensemble. This normalization step is key to determine if the increase in cavity mode intensity is due to better coupling between the defect ensemble and the cavity or is simply due to an increase in the number of photoluminescent defects in the ensemble.

*Room temperature spectra and cavity tuned spectra at 77 K*

PL spectroscopy of the nanobeam PCCs was taken at room temperature before and after processing steps. We chose cavities with resonance wavelengths that are blue shifted from the V1′ zero phonon line (ZPL) of the $V_{Si}^{-1}$. These same cavities were cooled to 77 K and gas-tuned into resonance with the V1′ ZPL at around 859 nm to measure Purcell enhancement before and after processing. The laser illumination and thermal annealing processes (Figure S1) result in increased intensity of the normalized fundamental cavity mode at room temperature, thus suggesting enhanced coupling of the $V_{Si}^{-1}$ defect ensemble to the PCC. Cavity tuning at 77 K confirms the improvement in cavity mode coupling to the $V_{Si}^{-1}$ defect via an observed increase in Purcell enhancement of the V1′ ZPL compared to the un-treated cavity (insets of Figure S1). Table S1 provides a summary of the change in background PL signal, change in normalized intensity of the fundamental cavity mode, and change in Purcell enhancement for nanobeams processed under different conditions.

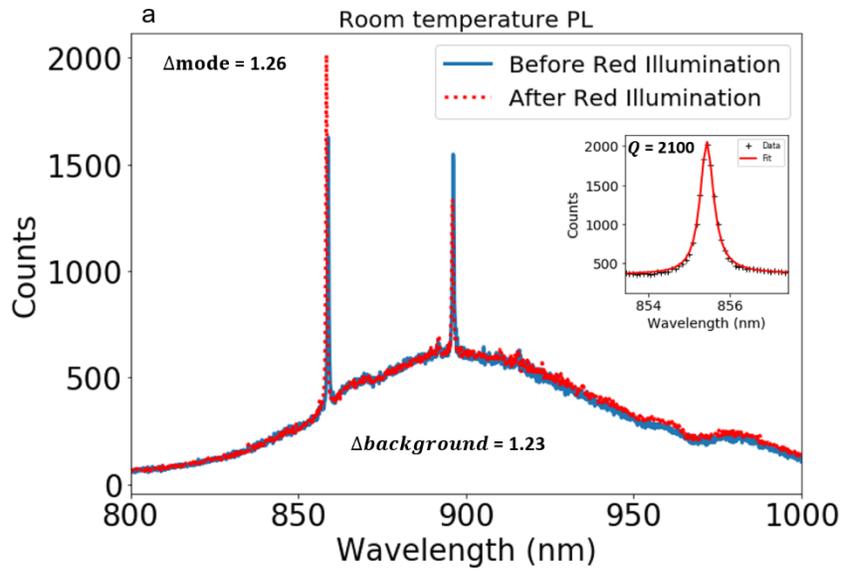
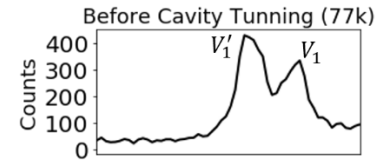
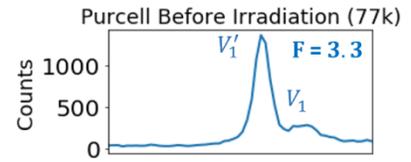
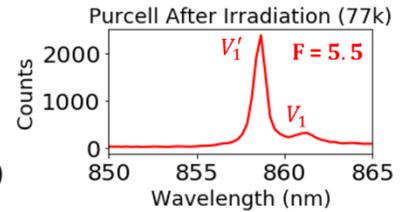
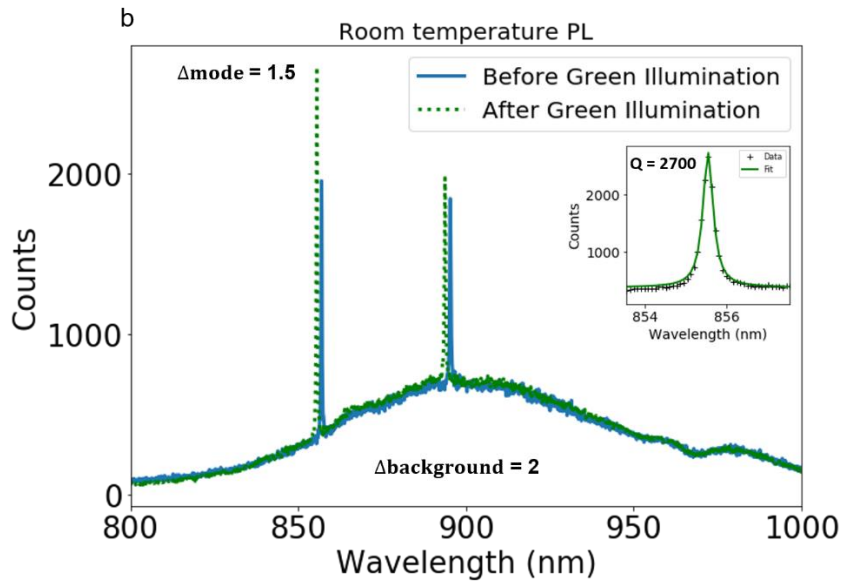
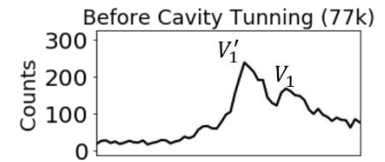
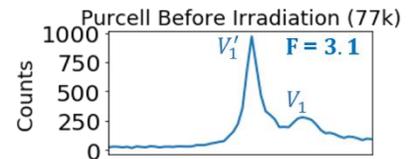
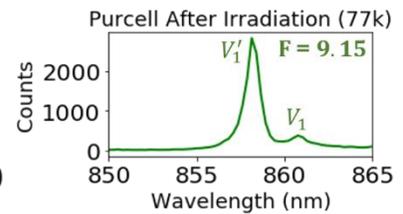

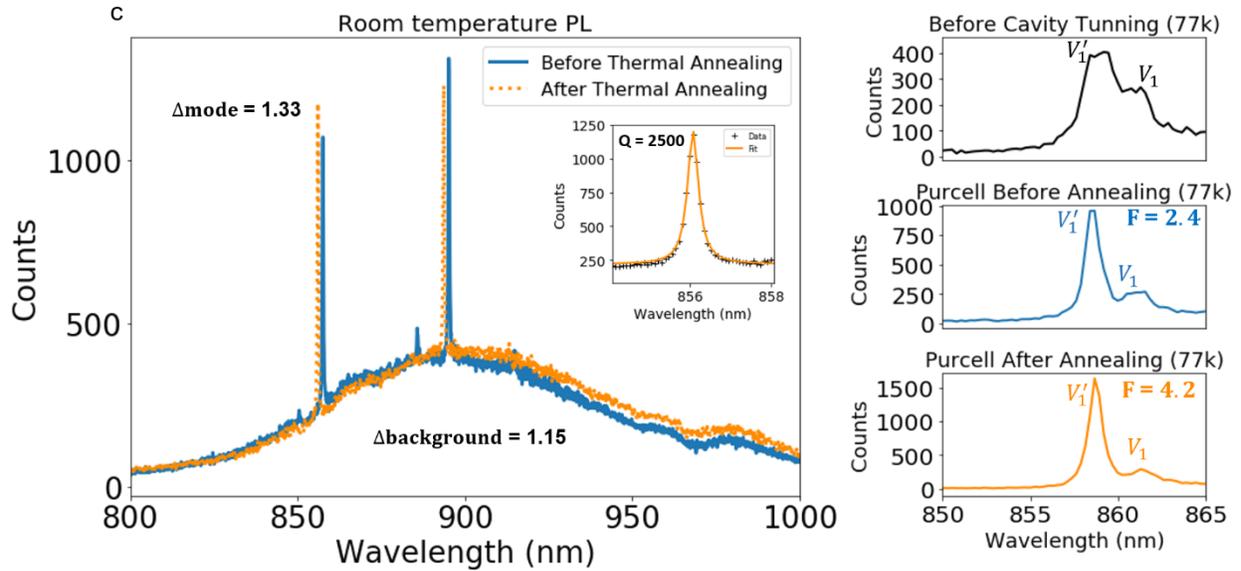

Figure S1. (a) (left) Normalized room temperature PCC spectrum before and after laser illumination, 785 nm at 50 mW for 6 hours, showing an increase in the normalized fundamental mode intensity at 858 nm by a factor of 1.26. The inset shows the spectra before and after laser illumination without background normalization. (right) 77 K cavity spectra (i) before tuning, (ii) after tuning for sample before laser illumination, (iii) after tuning for sample treated under laser illumination. Purcell enhancement is improved by a factor of 1.65. (b) (left) Normalized room temperature PCC spectrum before and after laser illumination, 532 nm at 30 mW for 6 hours, showing an increase in the normalized fundamental mode intensity at 855 nm by a factor of 1.5. The inset shows the spectra before and after laser illumination without background normalization. (right) 77 K cavity spectra (i) before tuning, (ii) after tuning for sample before laser illumination, (iii) after tuning for sample treated under laser illumination. Purcell enhancement is improved by a factor of 2.95. (c) (left) Normalized room temperature PCC spectrum before and after thermal annealing at 100 °C for 20 minutes, showing an increase in the normalized fundamental mode intensity at 857 nm by a factor of 1.33. The inset shows the spectra before and after laser illumination without background normalization. (right) 77 K cavity spectra (i) before tuning, (ii) after tuning for sample before laser illumination, (iii) after tuning for sample treated under thermal annealing. Purcell enhancement is improved by a factor of 1.75.

Table S1: Summary of PL changes due to laser illumination and thermal annealing, both at room temperature and when the nanobeams are cooled to 77 K and gas-tuned into resonance with the $V1'$ ZPL of the $V_{Si}^{-1}$.

| Operation | Change in background intensity | Change in normalized fundamental cavity mode intensity | Change in Purcell Factor for V1′ ZPL |
|---|---|---|---|
| Beam17 Red | 1.3 | 1 | 0.97 |
| Beam18 Red | 1.23 | 1.26 | 1.65 |
| Beam15 Green | 1.23 | 1.2 | 1.4 |
| Beam16 Green | 2 | 1.5 | 2.95 |
| Beam19 Thermal | 1.15 | 1.33 | 1.75 |

Modeling for likelihood of recombination of $C_i$ with adjacent $V_C$

We set up a simple system consisting of a single carbon interstitial next to a single carbon vacancy as a proxy for a single carbon interstitial next to a divacancy to assess the probability of recombination at 100 ºC for 20 minutes assuming the interstitial moves according to the barriers for the +1 charge state and that the vacancy moves according to barriers from earlier work.